\documentclass[11pt,fleqn,epsf]{article}
\usepackage{espcrc1}

\def\ee{\end{equation}}
\def\be{\begin{equation}}
\def\eea{\end{eqnarray}}
\def\bea{\begin{eqnarray}}
\def\ea{\end{eqnarray}}
\def\ba{\begin{eqnarray}}
\def\eeas{\end{eqnarray*}}
\def\beas{\begin{eqnarray*}}
\newcommand{\eqref}[1]{(\ref{#1})}
\usepackage{graphicx}
\title{\bf The multilevel pairing Hamiltonian versus the degenerate case}
\author{M.B. Barbaro~\footnote{Corresponding author: 
e-mail: barbaro@to.infn.it}
        \address[TO]{Dipartimento di Fisica Teorica, Universit\`a di Torino
and INFN, Sez. di Torino, Torino, Italy}, 
R. Cenni\address{Istituto Nazionale di Fisica Nucleare, 
Sez. di Genova, Genova, Italy },
S. Chiacchiera
\addressmark[TO],
A. Molinari
\addressmark[TO],
F. Palumbo
\address{INFN --- Laboratori Nazionali di Frascati, Frascati, Italy}
}
\begin{document}
\date{}
\maketitle
\abstract
\subsection*{Abstract}
{%
We study the pairing Hamiltonian in a set of non degenerate levels.
First, we review in the path integral framework 
the spontaneous breaking of the $U(1)$ symmetry occurring in such a system
for the degenerate situation.
Then the behaviors with the coupling constant of the ground state energy
in the multilevel and in the degenerate case are compared.
Next we discuss, in the multilevel case, an exact strong coupling expansion 
for the ground state energy which introduces the moments of the
single particle level distribution. 
The domain of validity of the expansion, which is known in the 
macroscopic limit, is explored for finite systems and its implications for the 
energy of the latter is discussed.
Finally the seniority and Gaudin excitations
of the pairing Hamiltonian are addressed and shown to display the same gap
in leading order.
}
\endabstract

{\small 
\noindent
{\em PACS:}\ 21.60.-n, 21.30 Fe, 24.10.Cn;
{\em Keywords:}\ Pairing interaction; Bosonization.}

\section{Introduction}

The pairing Hamiltonian
\be
\hat H= \sum_j \sum_{m=-j}^j e_{j} 
\hat \lambda^{\dag}_{jm} \hat \lambda_{jm} -g \sum_{i,j} 
\sqrt{\Omega_i\Omega_j} \hat A^{\dag}_i \hat A_j,
\label{HP}
\ee
where (note that this definition differs from the one of 
Refs.~\cite{Barbaro:2004nk,Barbaro:2003bb,Barbaro:2002fi} by a normalization
factor $1/\sqrt{\Omega_j}$)
\be
\hat A_j = \frac{1}{\sqrt{\Omega_j}} \sum_{m>0}(-1)^{j-m} 
\hat \lambda_{j,-m} \hat \lambda_{j,m}\ ,
\label{A}
\ee
represents a simplified version of the BCS model of superconductivity since it
assumes a constant two-body matrix element of the pairing force between all 
the single particle levels.

While BCS is generally applied to a macroscopic system, the
Hamiltonian \eqref{HP} is also suitable for dealing with finite systems.
Indeed it has been successfully applied to atomic nuclei where
$N$ fermions live in $L$ single particle levels, each with pair degeneracy 
$\Omega_j=j+1/2$. In this case 
in \eqref{HP} ${\hat \lambda}^{\dag}$, ${\hat \lambda}$ are the usual 
creation and annihilation fermion operators, 
$m$ is the third component of the angular momentum $j$, 
the $e_{j}$ are the single-particle energies (assumed to be $m$-independent) and 
$g$ is the strength of the pairing force.

One recovers BCS from \eqref{HP} by taking $\Omega_j=1$
for all the $L$ levels and letting $L$ and $N$ become very large
(as appropriate to a macroscopic system, e.g. electrons in a metal), but keeping
their ratio constant and $g L$ finite.

As is well-known 
in this limit the ground state of the grandcanonical version of \eqref{HP}
reads~\cite{Haag}
\be
|BCS> = \prod_{k>0} \left(u_k+v_k 
\hat \lambda^{\dag}_{{\vec k} \uparrow } 
\hat \lambda^{\dag}_{-{\vec k} \downarrow}
\right)|0> ~,
\label{BCS}
\ee
$|0>$ being the particle vacuum and
$u_k$, $v_k$ parameters taken to be real.
The operators in \eqref{BCS} create a pair of particles in time reversal
states.

The state \eqref{BCS} is not invariant under the group $U(1)$ of the global
gauge symmetry respected by \eqref{HP}, hence it is
associated with the spontaneous breaking of the particle
number conservation.

When applied to a system of a {\em finite}, fixed number 
$N$ of particles \eqref{BCS}
becomes an approximation, valid, however, to the order $1/N$.
This follows from the finding of Cambiaggio 
et al.~\cite{Cambiaggio:1997vz} that the Hamiltonian \eqref{HP} corresponds
to an integrable system: hence it is solvable (classically by quadrature,
according to the Liouville theorem).
Indeed Richardson~\cite{Rich} was able to find out the system of
algebraic equations yielding the exact solution which,
for a system made up of $n=N/2$ ($N$ even) pairs of fermions living in $L$ 
single particle
levels, reads~\cite{Barbaro:2003bb}
\be
|n> = C_n \prod_{\nu=1}^n {\hat B}^\dag_\nu |0>
\label{n}
\ee
being 
\be
{\hat B}_\nu = \sum_{j=1}^L \frac{1}{2 e_j-E_\nu} \hat A_j ~.
\label{B}
\ee
In \eqref{n} $C_n$ is a normalization constant and in \eqref{B} the quantities
$E_\nu$ ($\nu=1\cdots n$) are solutions of the following system of $n$
equations (the Richardson equations)
\be
\frac{1}{g} = \sum_{j=1}^L \frac{\Omega_j}{2 e_j-E_\nu} - 
\sum_{\mu(\neq \nu)=1}^n \frac{2}{E_\mu-E_\nu} ~,
\label{Rich}
\ee
whereas the $\hat A_j$ (defined in \eqref{A})
are referred to as hard-core boson operators and satisfy the commutation
relations
\be
\left[\hat A_j,\hat A_{j'}^\dag\right] = \delta_{jj'}\left(1-
\frac{\hat N_j}{\Omega_j}\right)~,
\ee
being
$
\hat N_j = \sum_{m=-j}^j \hat \lambda_{jm}^\dag\hat \lambda_{jm} 
$
the particle number operator of the $j$-level.

In Richardson's framework, the eigenvalues ${\cal E}$ of \eqref{HP} (in
particular the ground state energy) are obtained according to
\be
{\cal E}(n) = \sum_{\nu=1}^n E_\nu ~.
\ee
A question arises 
on the relationship between \eqref{n} and \eqref{BCS}
in the limit of $N$ and $L$ very large, the only situation where, strictly
speaking, the phase transition associated with
the spontaneous symmetry breaking occurs.
We shall later comment on this issue, already addressed by
Roman et al.~\cite{Roman:2002dh} who, in analyzing 
this limit, showed that the system of equations \eqref{Rich} becomes a non linear 
integral equation
and identified the order parameter associated with
the spontaneous breaking of  particle number conservation. This turns out
to be the BCS gap, namely
\be
\Delta = g \sum_{\mu>0} u_\mu v_\mu ~.
\label{Delta}
\ee
In general \eqref{Rich} can only be dealt with numerically.
However it is trivial to solve it when
only one single particle level of energy $\bar e$ and degeneracy $\Omega$
is available to the fermions.
In fact the system then reduces to the form 
\be
\frac{1}{g} = \frac{\Omega}{2{\bar e}-E_\nu} - 
\sum_{\mu(\ne\nu)=1}^n\frac{2}{E_\mu-E_\nu}
\label{Rich1La}
\ee
or
\be
E_\nu = 2{\bar e}-g \Omega
-2 g \sum_{\mu(\ne\nu)=1}^n\frac{E_\nu}{E_\mu-E_\nu} 
+4 g {\bar e} \sum_{\mu(\ne\nu)=1}^n\frac{1}{E_\mu-E_\nu}~,
\label{Rich1Lb}
\ee
which, upon summation over the index $\nu$, immediately yields the
well-known expression for the ground state energy
\be
{\cal E}=\sum_{\nu=1}^n E_\nu=2{\bar e}n-g n \left(\Omega-n+1\right) ~.
\label{Egs}
\ee
Note that in the above
the term quadratic in $n$ has a positive sign, 
reflecting the action of the Pauli principle.

In the $L=1$ case it is also easy to get 
the excitation spectrum of the Hamiltonian
\eqref{HP} which can only correspond to the breaking of pairs.
Indeed this occurrence can be formally accounted for by changing $\Omega$ into 
$\Omega-2 s$ and $n$ into $n-s$, 
$s$ (the so-called pair seniority) counting the number of broken 
pairs.
One ends up with the remarkably symmetric expression
\be
{\cal E}_s=\sum_{\nu=1}^n E_\nu=2{\bar e}n-g n \left(\Omega-n+1\right) 
+g s \left(\Omega-s+1\right) ~,
\label{Es}
\ee
which yields $2{\bar e}n$, as it should, for $s=n$, namely when all pairs are
broken, thus becoming blind to the pairing interaction.

\section{The path integral approach in the degenerate case}

The equations \eqref{Rich}, while exact, do not transparently convey the
occurrence of the spontaneous symmetry breaking taking place in the system.

To shed light on this point it turns out that a path integral approach,
while difficult, is more suited.
This scheme has been followed in Ref.~\cite{Barbaro:2004nk} starting from the 
discretized Euclidean action corresponding to the Hamiltonian \eqref{HP}, which in the
$L=1$ case reads~\cite{Negele}
\be
S=\tau\sum_{t=-N_0/2}^{N_0/2} \left\{
-g \Omega {\bar A}(t) A(t-1) +
\sum_{m=-j}^j \left[ {\bar\lambda_m}(t) \left( \nabla^+_t+{\bar e}\right)
\lambda_m(t-1) \right] \right\} ~,
\label{S}
\ee
$\tau$ being the time spacing, $N_0$ the number of points on the time 
lattice, the energy ${\bar e}$ is measured with respect to the chemical potential to 
select a sector of $N$ fermions
and $\lambda$, $\bar\lambda$ are odd Grassmann variables.
In terms of the latter
\be
A=\frac{1}{\sqrt{\Omega}}
\sum_{m>0} (-1)^{j-m} \lambda_{-m}\lambda_{m}~,
\ee
\be
{\bar A}=\frac{1}{\sqrt{\Omega}}
\sum_{m>0} (-1)^{j-m} {\bar\lambda_{m}}{\bar\lambda_{-m}}
\ee
and the discretized time derivative is defined as follows
\be
\left( \nabla^\pm_t f\right)(t) = \pm\frac{1}{\tau}
\left[f(t\pm 1)-f(t)\right] ~.
\ee
Next the action \eqref{S} is dealt with through the Hubbard-Stratonovitch
transformation by introducing a bosonic field $\eta$ cast in the
polar representation:
\be
\eta = \sqrt{\rho} e^{2 i \theta}
\ ,\ \ \ \ \ \ \ \
{\bar\eta} = \sqrt{\rho} e^{-2 i \theta} ~.
\label{eta}
\ee
In the end the following effective action, equivalent to \eqref{S}
and describing the same physics of the Hamiltonian \eqref{HP},
\be
S_{\rm eff} = \tau\sum_t g\rho - {\rm Tr} \ln \left(-q^- q^+ + g^2\rho\right)
\label{Seff}
\ee
is obtained.
In \eqref{Seff} the trace is taken both on the time and on the quantum number $m$
and
\be
q^\pm = e^{\mp i \theta} \nabla^{\pm}_t e^{\pm i \theta} \pm {\bar e} ~.
\ee
Now in the above we see that the field $\theta$
\begin{itemize}
\item[a)] appears in the action only under derivative,
\item[b)] lives in the coset space of the broken group $U(1)$ with
respect to the unbroken group $Z_2$.
This follows from the requirement of making the change of variable
\eqref{eta} to be one to one. Indeed for this to occur it must be
$0\leq\theta<\pi$.
\item[c)]Moreover, at variance with the Hamiltonian \eqref{HP}, the $U(1)$
symmetry
is now non-linearly realized in the invariance of the action \eqref{Seff}
under the substitution
$
\theta\to\theta+\alpha ~,
$
 $\alpha$ being time independent.
\end{itemize}
The above features characterize a Goldstone field (see \cite{Weinberg}).

These findings show that in the degenerate case the system lives in a regime 
where the $U(1)$ symmetry
is spontaneously broken not only at the macroscopic, but at the microscopic
scale as well.

According to the Goldstone theorem when a continuous symmetry is broken
a bosonic field appears with a dispersion relation which vanishes in 
the thermodynamic limit.

Of course
in a finite system, like a deformed nucleus, where the $O(3)$ symmetry 
is spontaneously broken , the Goldstone boson shows up as a rotational 
band whose frequency is not vanishing, but, however, 
tends to zero as the system becomes larger and larger.
Likewise in the pairing problem the $U(1)$ symmetry is broken and in the 
degenerate case the energy of the associated Goldstone mode depends upon 
the pair number $n$ according to \eqref{Egs}.
Actually setting (see \cite{Barbaro:2004nk})
\be
\nu=n-[n_0] ~,
\ee
where the square bracket means integer part and
\be
n_0 = \frac{\Omega+1}{2}-\frac{\bar e}{g} 
\ee
is the value of $n$ where \eqref{Egs} reaches its minimum, 
one sees that the energy of the Goldstone mode 
goes like $g$, whereas the energy associated with the variable $s$, hence
corresponding to the breaking of pairs, goes like $g\Omega$, thus lying
at a much higher energy.
Microscopically the Goldstone mode
corresponds to the addition or to the removal of a pair of fermions.

In the path integral framework another point is of relevance.
In dealing with $S_{\rm eff}$ with the saddle point expansion one starts by 
searching for its minimum at constant fields.
It turns then out that {\em the minimum of the effective action} occurs for 
$\rho=\bar\rho$, being~\cite{Barbaro:2005tp}
\be
{\bar\rho} = \frac{1}{(2 g)^2} \left[
(g\Omega)^2-4{\bar e}^2\right] = \frac{\Delta^2}{g^2} ~,
\label{rhobar}
\ee
where $\Delta$, see \eqref{Delta}, is the well-known gap characterizing the BCS
theory in the $L=1$ case. Thus $\bar\rho$, but for a factor which
renders it dimensionless, coincides with the gap.
Furthermore, choosing $\bar e=0$, \eqref{rhobar} yields $g\Omega=2\Delta$.

\section{The Richardson approach in the multilevel case}

We now study the case when $L$ single particle levels are active.
Our aim here is 
to ascertain how their presence 
modifies the previous results obtained in the degenerate case, 
specifically whether the system still lives in the phase where
the $U(1)$ symmetry is spontaneously broken.
In this connection it helps, as we shall see,
to derive an
exact analytic expansion of the lowest eigenvalue of \eqref{HP} in the
inverse powers of the coupling constant $g$. 
This item has been lately pursued by Yuzbashyan et al.~\cite{Yuz} in the
framework of the Richardson equations.
We also mention that this 
problem has been recently tackled in the path 
integral formalism as well~\cite{Pal}.
We shall do the same here, but
providing an expression for the expansion coefficients 
valid for \emph{any} single particle level (s.p.l.)
distribution and
more transparently
linked to the moments characterizing the latter.
Moreover new results concerning
${\overline g}$, namely the value of the coupling constant defining the domain 
where the expansion holds, will be presented.

\begin{figure}[t]
\vspace{-2cm}
\begin{minipage}[b]{0.5\textwidth}
\centering
\includegraphics[scale=0.7,clip,angle=0]{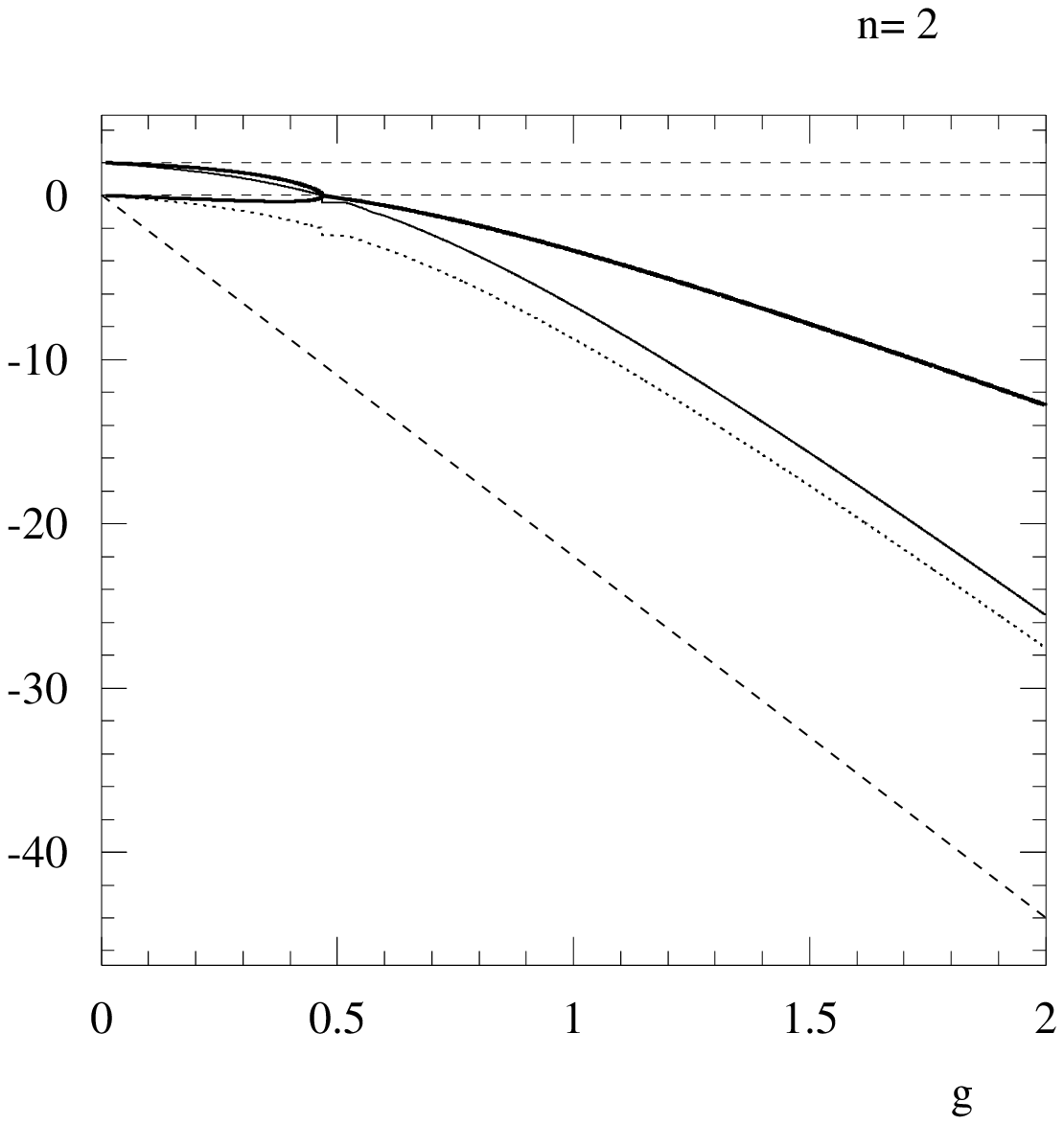}%
\end{minipage}
\begin{minipage}[b]{0.5\textwidth}
\centering
\includegraphics[scale=0.7,clip,angle=0]{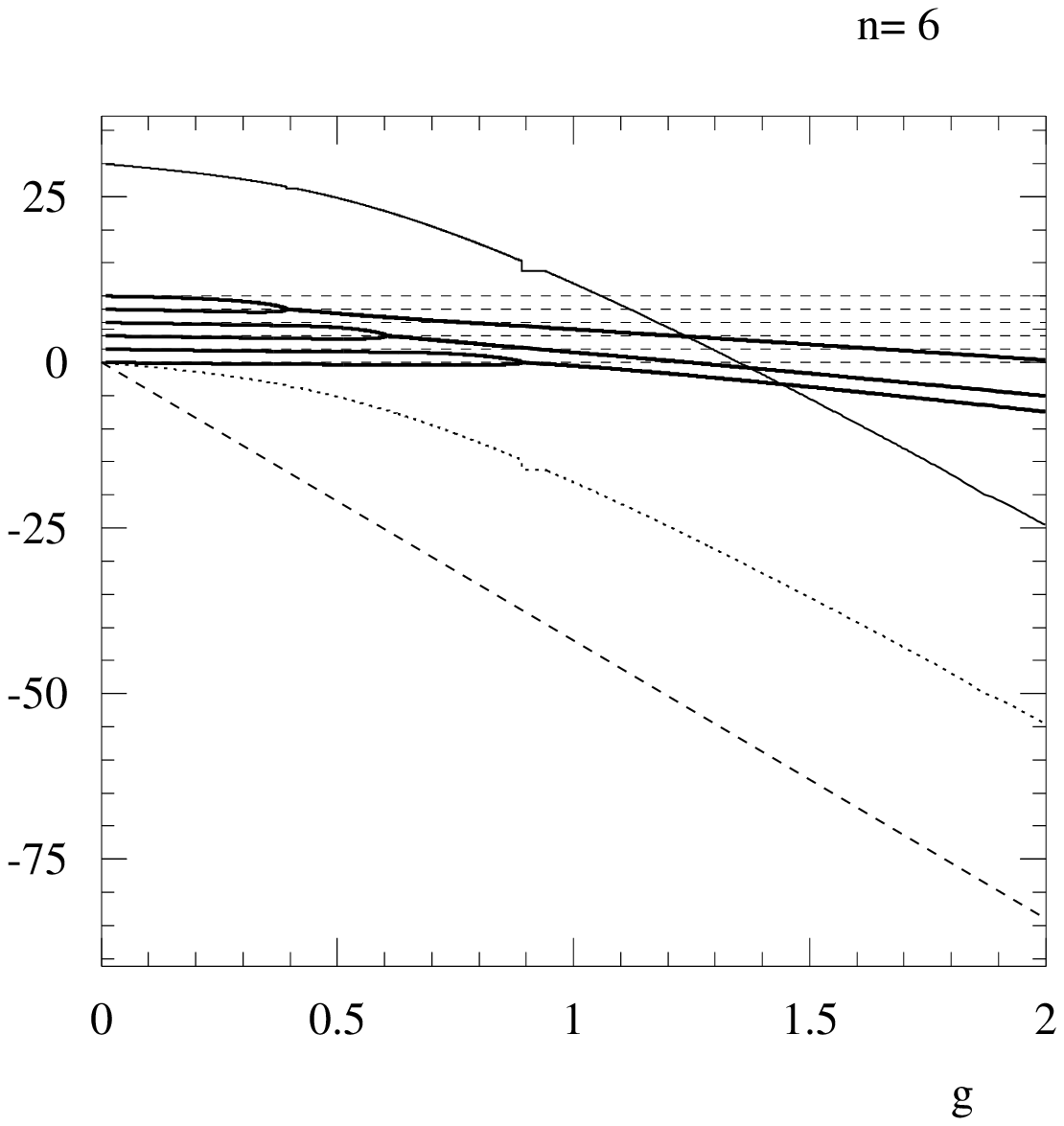}%
\end{minipage}
\begin{minipage}[b]{0.5\textwidth}
\centering
\includegraphics[scale=0.7,clip,angle=0]{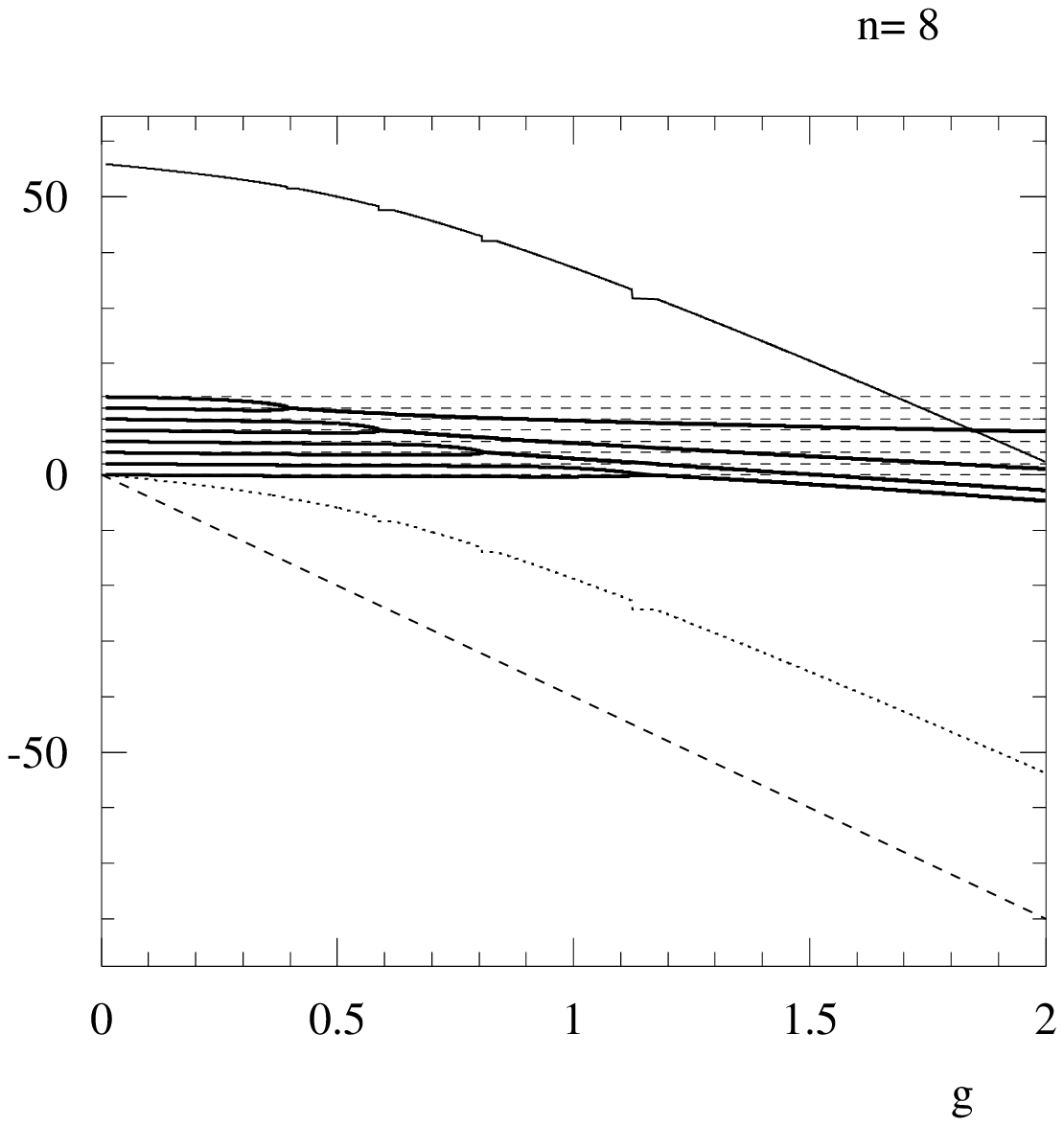}%
\end{minipage}
\begin{minipage}[b]{0.5\textwidth}
\centering
\includegraphics[scale=0.7,clip,angle=0]{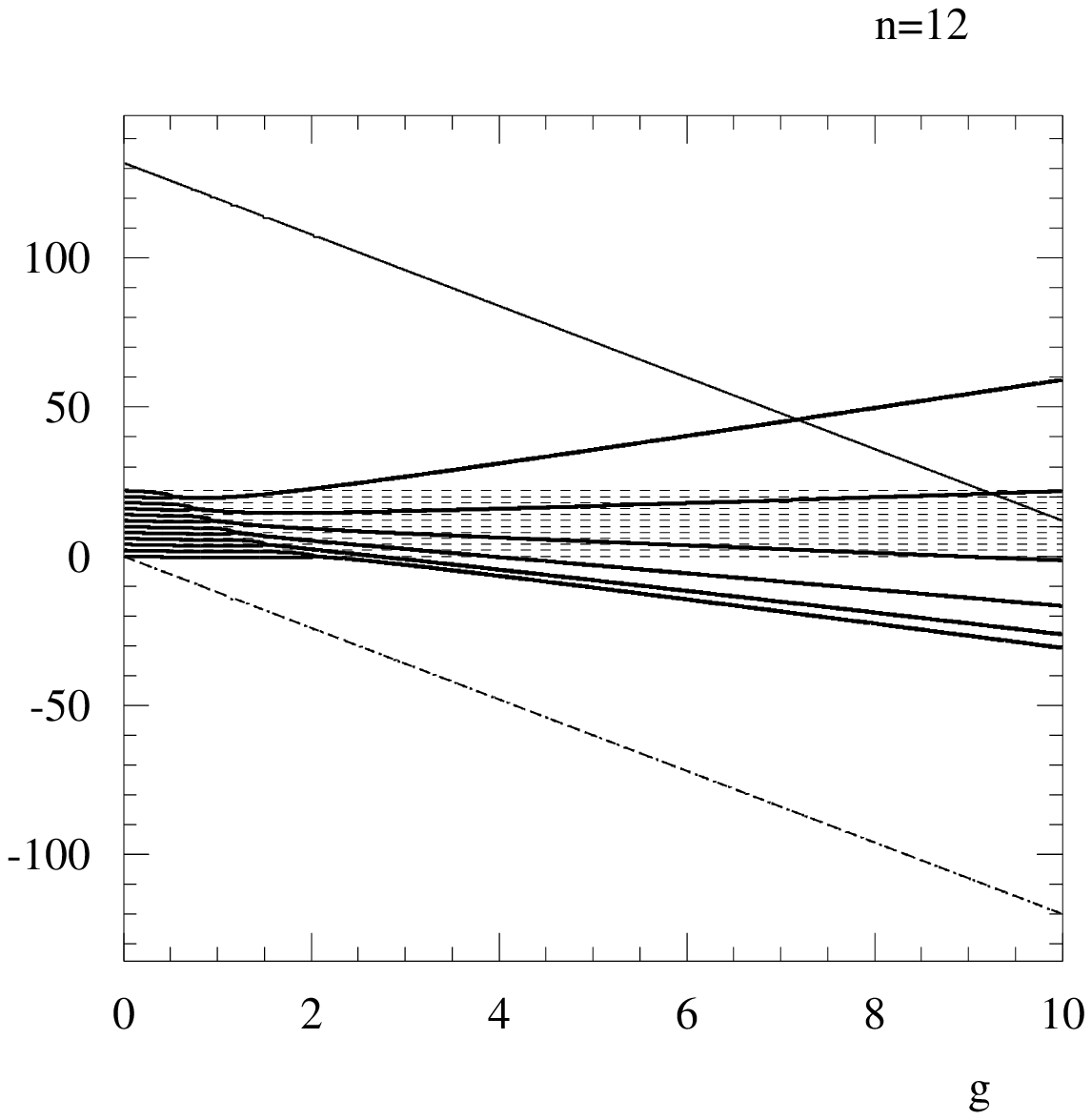}%
\end{minipage}
\vspace{-3cm}
{\caption{
The real part of the exact ground state solutions $E_\nu$ (thick solid curves) 
of the Richardson's equations (6) for a 
system of $n$=2,6,8 and 12 pairs living in a set of 
L=12 equispaced non-degenerate ($\Omega_k=1$) single particle levels with
energies $e_k=k$ ($k=0,\cdots 11$) are displayed versus 
the strength $g$ together with their sum ${\cal E}(n)$
(thin solid curves); the dotted curves represent the shifted total energy
${\cal E}(n)-2\sum_{k=1}^n e_k$ and the dashed curves the energy of $n$ 
pairs living on one level of energy $e=0$ and degeneracy $L$.
The horizontal dashed lines represent the unperturbed solutions and 
the cuspids are due to numerical imprecision close to the critical points.
These could be avoided by employing the Rombouts method \cite{Rom}.
}\label{fig1}} 
\end{figure}

\subsection{Comparison with the degenerate case}
We consider the $g$-dependence of the ground state energy to 
assess how it is affected by the presence of the $L$ single particle levels. 
For sake of illustration in this Subsection 
we confine ourselves to explore a model of equispaced
single 
particle levels with unit energy distance and
with unit pair degeneracy ($\Omega_j=1$). In Fig.~\ref{fig1} we display the 
Richardson solution for $L=\Omega=12$, for $n$=2, 6 (half filling), 8, 
12 (full filling). The energy of the degenerate case \eqref{Egs} is also 
shown as a reference.
Moreover the $g$-behavior of the energy of each pair is displayed.

It is clearly apparent in
Fig.~\ref{fig1} the merging mechanism which allows the escaping of the
 pair energies from the grid of the single particle energies from
where they originate at $g=0$ (see
Refs.~\cite{Roman:2002dh,Barbaro:2003bb}).
In fact, considering the ground state, the pair energies start out, 
from the lowest bare single particle energies, being real at small $g$ 
and then, two by two,
{\em merge} at the energy of a single particle level (identified by the
superscript $i$) for critical values of the coupling $g_{\rm crit}^i$:
clearly $[n/2]$ of these exist. 
For $g>g_{\rm crit}^i$ the two pair energies associated
with the index $i$ become complex conjugate and, what is important,
their common real part behaves almost {\em linearly in $g$}.
Thus when  $g>g_{\rm crit}^{\rm max}$ all the pair energies become complex 
with real parts displaying a behavior close to linear in $g$ and, 
remarkably, they add up to yield a 
downwards pointing straight line with the same slope as \eqref{Egs}, 
but upwardly shifted with respect to this one because of the contribution 
stemming from the single particle energies, which is obviously 
$g$-independent. 
One can accordingly argue that when $L$ single particle levels are active, 
the system, at variance with the degenerate case, can live in different 
regimes.
These in turn depend upon the filling of the levels.
Here we consider the half filling situation where 
the system can live in three different regimes 
according to whether  $g<g_{\rm crit}^{\rm min}$ (normal fermionic regime, where the
wave function of each pair has essentially only one component), 
$g>g_{\rm crit}^{\rm max}$ (superfluid regime, where
the wave function of each pair is spread out over all the $L$ single particle 
unperturbed levels) 
and $g_{\rm crit}^{\rm min}<g<g_{\rm crit}^{\rm max}$ (mixed regime).

From the figure it also appears that 
the energy of the degenerate and non-degenerate case are markedly
different only for $g<g_{\rm crit}^{\rm max}$.
However this difference becomes less and 
less pronounced as the ratio $n/\Omega$ approaches unity (full filling).
Indeed for $g<g_{\rm crit}^{\rm max}$, but $n$ 
sufficiently large, the non-linear behaviors of the exact $E_\nu(g)$ 
cancel out leading to a system's total energy linear in $g$ as in 
the degenerate case. 
This occurrence relates to the absence of an excitation spectrum 
in the full filling situation in both the degenerate and non-degenerate cases.

It is interesting to check whether the above findings agree with the Anderson
criterion~\cite{Anderson} for establishing how small a superconductor can be.
This states that in the superconducting phase it must be
\be
\frac{\Delta}{d} > 1~,
\label{And}
\ee
$d$ being the average distance between the levels of the finite system and
$\Delta$ the gap.
Now by combining the above with \eqref{rhobar} it follows that a 
superconducting regime sets in for
\be
g>g_{\rm crit}^{\rm max} = \frac{2}{\Omega} \sqrt{{\bar e}^2+d^2}~,
\label{ggcr}
\ee
which indeed appears well fulfilled in our numerical analysis
(which assumes $d=1$, $\Omega=L$ and $\bar e$ given by \eqref{mom}).
Hence \eqref{And} holds valid in our model.

Finally a comment on the above recalled escaping mechanism, responsible for
the setting up of a strongly collective ground state, is in order.

Indeed the way the pair energies evade the grid 
changes its nature when the number of pairs substantially exceeds the 
half filling of the single particle levels. This is
clearly apparent in the last panel of Fig.~\ref{fig1} (note that there a larger
span of $g$ is
displayed), where the energies of the four upper pairs are seen to grow rather 
than to decrease with $g$.

In fact the pair energies are differently affected by the pairing
force. In particular the lowest lying pairs feel an attractive interaction
whereas the upper lying ones feel a repulsive force. Since the sum of the pair 
energies is constrained to yield the degenerate result, it is not surprising 
that when $n$ is large the upper lying 
pairs contribute to the total system's energy positively and not negatively.
In other words the linear behavior in $g$ of their energies 
has a positive (and not a negative) slope.

\subsection{The expansion of the ground state energy}

We now turn to derive an exact analytic expression for the 
$1/g$ expansion for the system ground state energy 
when $L$ s.p.l. and $n$ pairs are active. 
This we achieve along the lines illustrated in Ref.~\cite{Yuz}, but 
we prefer to express the coefficients of 
the expansion in terms of the mean energy 
\be \label{emedia}
\bar e=\frac{1}{\Omega}\sum_{\mu=1}^L \Omega_\mu e_\mu
\ee
and of the moments
\be\label{mom}
m^{(k)} = \frac{1}{\Omega} \sum_{\mu=1}^L \Omega_\mu(e_\mu-{\bar e})^k
\ee
of the s.p.l. distribution
(the first of these, namely when
$k=2,3$ and $4$, correspond to the variance, skewness and kurtosis).
In the above $\Omega=\sum_{\mu=1}^L\Omega_\mu$ is the total pair degeneracy. 
This allows to better grasp the impact of s.p.l. 
on the spectrum of the system.

We obtain 
\begin{equation}
{\cal E}_{gs}(n) =
2 n {\bar e} 
-g n (\Omega-n+1)
- d \sum_{j=1}^\infty \alpha_j \left(\frac{d}{g}\right)^{j}.
\label{expansion}
\end{equation}
In (\ref{expansion})  $d$ is the average distance between the single particle levels and
\be
\alpha_j=\Omega\sum_{p=0}^j 2^{p+1} \ a_{p+1}^{j-p} 
\left\{
m^{(p+1)}+\sum_{l=0}^p {p+1\choose l} {\bar e}^{p+1-l} m^{(l)}
\right\}~.
\label{alpha}
\ee
The coefficients $a$ in \eqref{alpha} are related through the following
recurrence relations
\be
a_{p+1}^{q} = -\frac{1}{\Omega-p}
\left\{
\frac{a_p^{q}}{d}+
(1-\delta_{p0}) \sum_{k=1}^p \sum_{s=0}^{q} a_{p+1-k}^{q-s} \ a_k^s
+\Omega (1-\delta_{q0}) \sum_{k=1}^{q} 2^k \ a_{p+k+1}^{q-k}\  
\sum_{l=0}^k {k\choose l} {\bar e}^{k-l} m^{(l)}
\right\}~,
\label{recurrence}
\ee
which allows to evaluate the expansion \eqref{expansion} at any 
given order starting from $a_0^k=n \delta_{k0}$. 

Eq.~\eqref{expansion} represents, when it converges, 
the Laurent expansion of
${\cal E}_{gs}$ around the simple pole it displays at $g=\infty$.
The first two terms of the expansion just correspond to the ground state
energy of the degenerate case.
The other terms thus account for the role of the s.p.l. distribution: indeed
the $n$-th order in $1/g$ involves the moments of the s.p.l. 
distribution up to the order $n+1$. 
Clearly the smaller $g$ is, the more moments are needed to get a faithful
representation of ${\cal E}_{gs}(n)$.
This is supported by all the cases we have numerically explored.

Keeping the first five terms only one gets (note that our 
term $1/g^2$ differs from the one in Ref.~\cite{Yuz})
\ba
{\cal E}_{gs}(n) &=&
2 n {\bar e} 
-g n (\Omega-n+1)
-2\frac{m^{(2)}}{(g\Omega/2)}\times\frac{n(\Omega-n)}{\Omega-1}
+2\frac{m^{(3)}}{(g\Omega/2)^2}\times
\frac{n(\Omega-n)(\Omega-2n)}{(\Omega-1)(\Omega-2)}
\nonumber\\
&-&\frac{2}{(g\Omega/2)^3}\times
\frac{n(\Omega-n)}{(\Omega-1)^2(\Omega-2)(\Omega-3)}
\left\{
\left[\Omega^2(\Omega-1)-n(\Omega-n)(5\Omega-6)\right] m^{(4)} 
\right.
\nonumber\\
&-&\left.\left[\Omega^2(2\Omega-3)-n(\Omega-n)
\frac{3(\Omega-1)(3\Omega-4)-\Omega}{\Omega-1}\right] \left[m^{(2)}\right]^2
\right\}.
\label{EgsY}
\ea

For one pair \eqref{EgsY} yields back the result of
Ref.~\cite{Barbaro:2002fi}, whereas when all the levels are filled
($n=\Omega$) it yields back \eqref{Egs},
characterized by
the linear $g$-behavior 
displayed in Fig.~\ref{fig1}. 
Moreover, when the levels are equispaced 
only even moments, hence odd powers of $1/g$, enter into the expansion.

\begin{figure}[hbt]
\hspace{-2cm}
\begin{minipage}[t]{0.5\textwidth}
\centering
\includegraphics[scale=0.7,clip,angle=0]{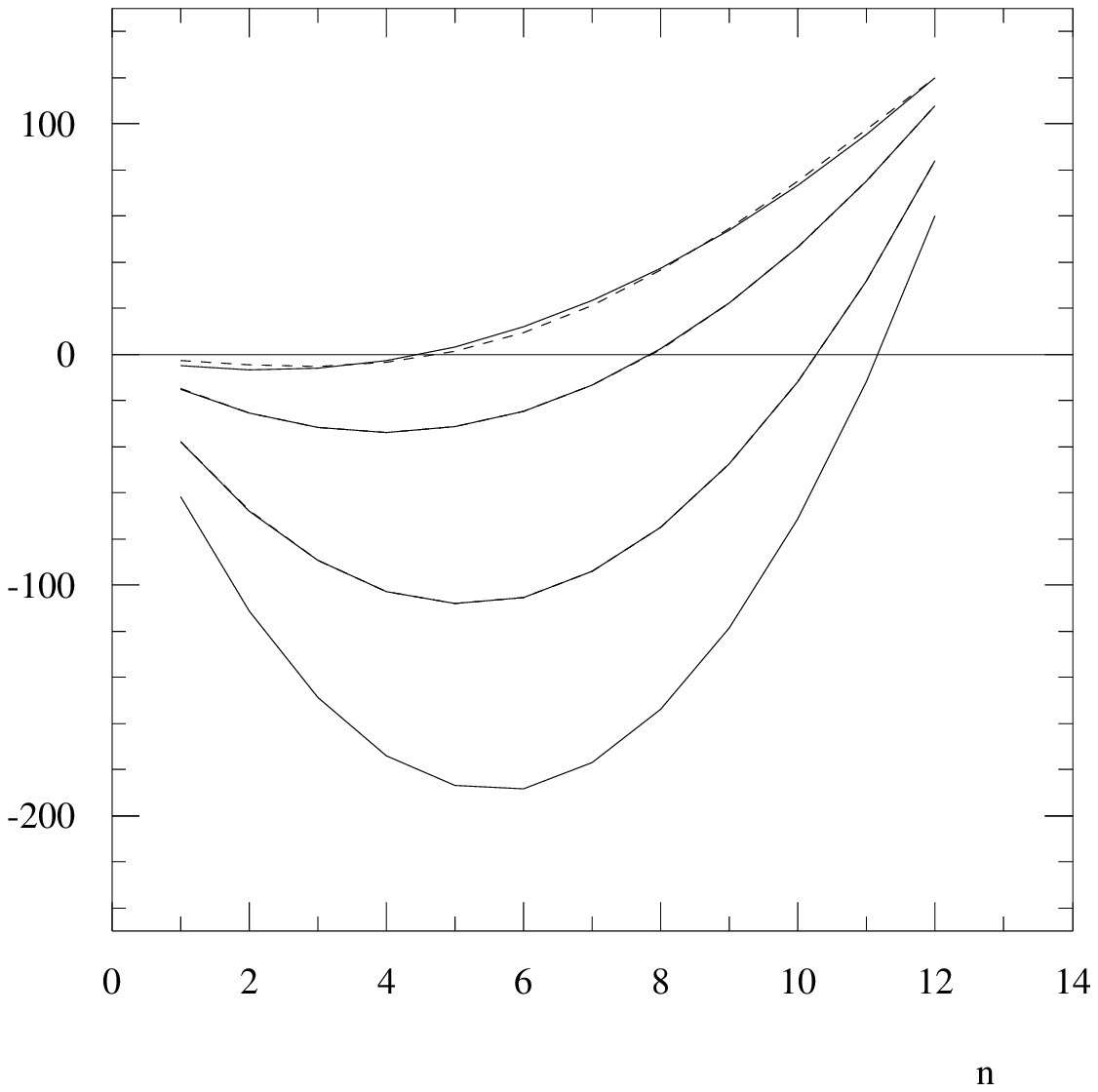}%
\end{minipage}
\begin{minipage}[t]{0.5\textwidth}
\includegraphics[scale=0.7,clip,angle=0]{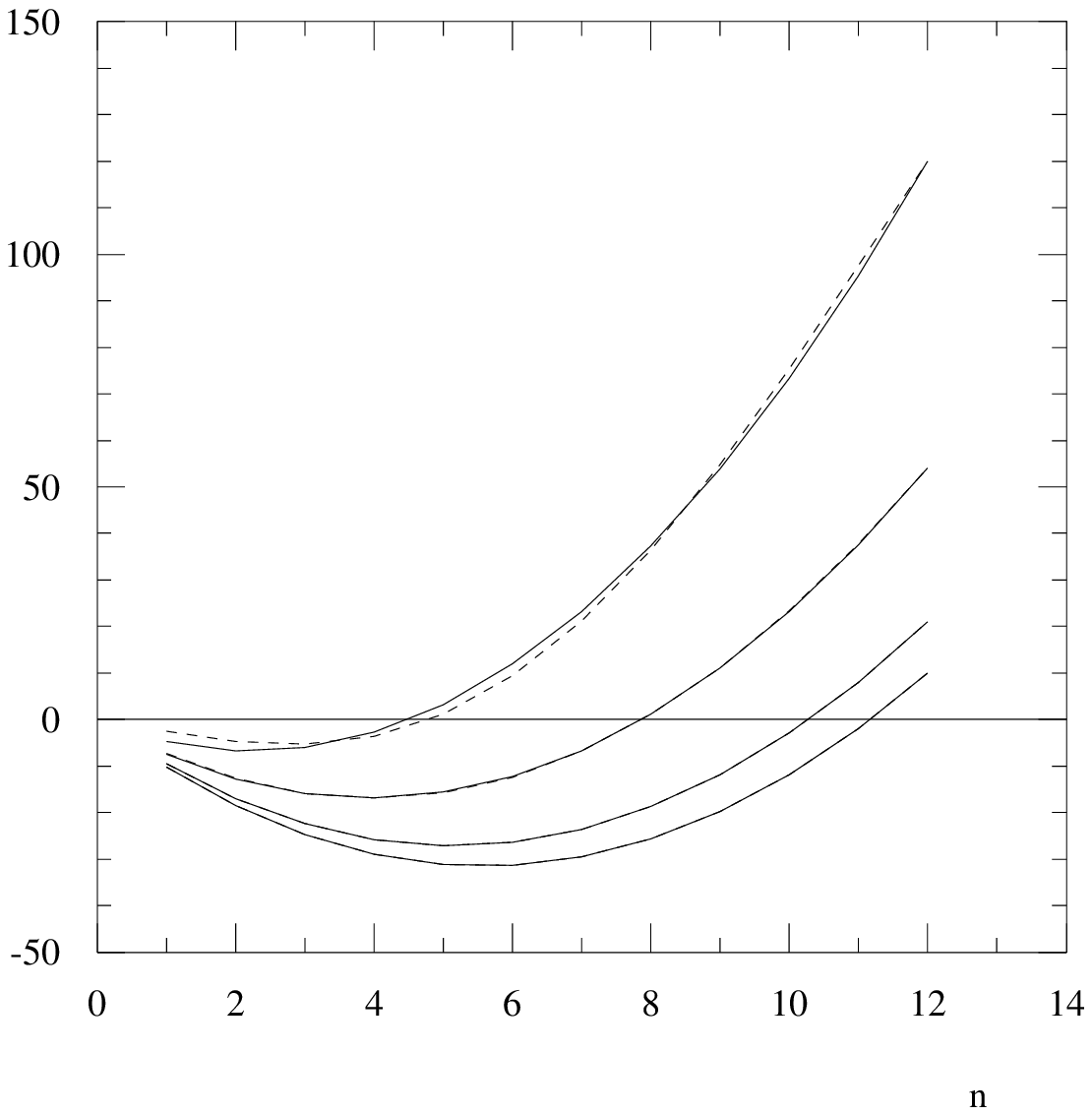}
\end{minipage}
\vspace{-1cm}
{\caption{Left panel: the exact (solid curves) ground state energy for the same model 
as in Fig.~\ref{fig1} is displayed versus the pair number $n$ for $g$=1 (upper curve), 
2, 4 and 6 (lower curve), respectively; 
the dashed curves (visible only in the $g$=1 case) correspond to the 
approximate solution of Eq.~(29). Right panel: the same curves divided by $g$;
for very large $g$ the curves tend to the parabola $-n(L-n+1)$.
}\label{fig2}} 
\end{figure}

In Fig.~\ref{fig2}, we compare the $n$-behavior of the ground state energy of 
the system as obtained in the Richardson's framework and from \eqref{EgsY}.
They are both displayed for a few values of $g$ assuming 
$L=12$ and the same set of equispaced single particle energies as before.
From these results one can gauge the validity of the expansion \eqref{expansion} 
even when only a few terms of the latter are kept.
It is 
gratifying that few moments of the level distribution are enough to get 
a good representation of the true solution, providing $g$ is not too close to
the boundary of the domain of convergence of the expansion.

Accordingly we address the problem of establishing 
the domain of convergence of the expansion 
\eqref{expansion}. To explore this issue we first 
consider the half filling situation in the following
limiting cases:
the macroscopic limit and when only one pair
lives in two levels. 
In the former
we assume that all the unperturbed levels have unit pair degeneracy 
and live in 
a finite energy band of width $2\hbar\omega_0=2dn=d\Omega$, in such a way that
when $g\to 0$ the ratio $g/d$ remains finite.
We then obtain for the condensation energy
\ba
{\cal E}_{gs}^{cond}(n) &=&{\cal E}_{gs}(n)-2n\overline e=
-n\hbar\omega_0\left\{\frac{g}{d}+\frac{1}{3}\frac{d}{g}-\frac{1}{45}\left(\frac{d}{g}\right)^3
+\cdots \right\}
\label{EgsY_TL}
\ea
which coincides, when it converges, namely for 
\be
\frac{g}{d} > \frac{\overline g}{d} = 1/\pi~, 
\ee
with the Laurent $d/g$-expansion 
of the BCS energy
\ba
{\cal E}_{BCS}^{cond}(n) &=&
-n\hbar\omega_0\coth\left(\frac{d}{g}\right)
\label{EgsY_BCS}
\ea 
in accord with
the long known result that the BCS energy cannot be
computed perturbatively for small $g/d$.

On the other hand the macroscopic limit of the Richardson solution
has been explored in Ref.~\cite{Roman:2002dh}, where it is shown that
in this instance
\ba
 \left(\frac{g}{d}\right)^{\rm max}_{\rm crit}=\frac{1}{\rm{arcsinh}(1)}=1.1345
\qquad\textrm{and}\qquad\left(\frac{g}{d}\right)^{\rm min}_{\rm crit}=0.
\label{gmax}
\ea

Thus the domain of validity of the expansion \eqref{EgsY_TL} 
does not directly relate to the critical values of the coupling constant:
indeed although these correspond to branch points  of the pair energies 
\cite{Rich} these singularities cancel out in the expression
for the system's global energy.

Turning to the case of one pair living in two levels
with unit degeneracy and having an energy distance $d$ from 
Eq.~(\ref{EgsY}) one gets
\be \label{eqa}
{\cal E}=-g -g\left(1+\frac{1}{2}\left(\frac{d}{g}\right)^2-\frac{1}{8}\left(\frac{d}{g}\right)^4
+\frac{1}{16}\left(\frac{d}{g}\right)^6+\cdots\right)
\ee
which coincides, when it converges, namely for
\be \label{eqb}
\frac{g}{d} > \frac{\overline g}{d} = 1~, 
\ee
with the well-known result 
\be\label{eqc}
{\cal E}=-g-\sqrt{d^2+g^2}~.
\ee
Of course the range (\ref{eqb}) is not related to critical values
of $g$ which are now non existent. Note that this most simple case 
already shows that a $\overline g$ can exist without any $g_{\rm{crit}}$.
This occurrence goes in parallel with the findings of \cite{Rin} where 
the same is seen to happen in a much more involved situation.

Thus for both systems above examined ($n=1$ and $n=\infty$ at 
half filling) the function ${\cal E}_{gs}(g)$ displays a singularity
(either a branch point or a pole)
on the imaginary axis of the complex $g$-plane. By contrast
the branch points associated with the $g_{\rm crit}$ all lie
on the real axis.

To gain insight on the actual value of $\frac{\overline g}{d}$ 
in a generic system we quote 
in Tables $1$ and $2$ the values of 
$\left(\frac{g}{d}\right)_{\rm crit}^{\rm min}$,
$\left(\frac{g}{d}\right)_{\rm crit}^{\rm max}$ 
and
of 
$\frac{\overline g}{d}$ for a few values of $n$.
The radius of convergence $\frac{\overline g}{d}$ 
of the expansion has been computed using
the Cauchy-Hadamard criterion
\be
\frac{{\overline g}}{d} 
= \lim_{j\to\infty}|\alpha_j|^{1/j}~.
\ee
To assess the sensitivity to $j$ of the above in Table $2$
we report in the first column the results obtained with $j=21$
and in the second column those obtained  (at least in a few cases)
with $j=199$.

From our results it follows that 
the radius of convergence $\frac{\overline g}{d}$ of
the expansion \eqref{expansion} for any $n$ appears to occur, 
for the half-filling case and for equispaced s.p.l. of unit pair
degeneracy,
in the domain ranging from 
$\left(\frac{g}{d}\right)_{\rm crit}^{\rm min}$
to
$\left(\frac{g}{d}\right)_{\rm crit}^{\rm max}$.
This interval increases with the pair number $n$, 
as it can be inferred from Table 1, where
$\left(\frac{g}{d}\right)_{\rm crit}^{\rm max}$
($\left(\frac{g}{d}\right)_{\rm crit}^{\rm min}$)
is seen to increase (decrease) with $n$.

We conclude that $\frac{{\overline g}}{d}$ 
\begin{itemize}
\item[a)] varies in the range $\frac{1}{\pi}\leq\frac{{\overline g}}{d}\leq 1$,
for any $n$,
\item[b)] decreases with $n$,
\item[c)] lies, for a given $n$, in the range
$
\left(\frac{g}{d}\right)^{\rm min}_{\rm crit}
\leq\frac{{\overline g}}{d}\leq 
\left(\frac{g}{d}\right)^{\rm max}_{\rm crit}
$ .
\end{itemize}
Note that $\frac{{\overline g}}{d}$
, even for $n=500$, is still far from $1/\pi$
: hence the macroscopic limit of the pairing Hamiltonian
appears to be reached very slowly.

\begin{table}[ht]
\begin{center}
\begin{tabular}{||c|c|c||}
\hline\hline  \raisebox{0pt}[14pt]{} \raisebox{-7pt}[16pt]{} 
$n$ & $\left(\frac{g}{d}\right)^{\rm min}_{\rm crit}$
& $\left(\frac{g}{d}\right)^{\rm max}_{\rm crit}$
\\
\hline\hline
    2  & 0.66 & - \\ 
    4  & 0.46 & 0.82 \\ 
    6  & 0.40 & 0.90 \\ 
    8  & 0.35 & 0.94 \\ 
    10 & 0.32 & 0.96 \\ 
    20 & 0.27 & 1.03 \\
 $\infty$ & 0 & 1.13
\\ \hline\hline 
\end{tabular} 
\caption{For a few values of the pair number $n$ the largest and smallest
critical values of the coupling constant are quoted at half filling.
The number of critical points is $n/2$ and clearly for $n=2$ only one
of them exists.
}
\label{tab:1}
\end{center} 
\end{table} 

\begin{table}[ht]
\begin{center}
\begin{tabular}{||c|c|c||}
\hline\hline  \raisebox{0pt}[14pt]{} \raisebox{-7pt}[16pt]{} 
$n$ & $|\alpha_{21}|^{1/21}$&  $|\alpha_{199}|^{1/199}$
\\
\hline\hline
    1  & 0.82 & 0.96\\
    2  & 0.69 &\\ 
    50 & 0.41&\\
    100& 0.39 & 0.65\\
    500& 0.36 & 0.64\\    
 $\infty$& 0.31&$0.318\simeq1/\pi$
\\ \hline\hline 
\end{tabular} 
\caption{The radius of convergence of the expansion \eqref{expansion} 
for various
pair numbers $n$ at half-filling and equispaced single particle levels
of unit pair degeneracy.
}
\label{tab:2}
\end{center} 
\end{table}

\begin{table}[ht]
\begin{center}
\begin{tabular}{||c|c||}
\hline\hline  \raisebox{0pt}[14pt]{} \raisebox{-7pt}[16pt]{} 
$n$ & $|\alpha_{199}|^{1/199}$
\\
\hline\hline
    1   & 0.0096 \\
    50  & 0.0077\\ 
    100 & 0.0068\\
    150 & 0.0077\\
    199 & 0.0096\\    
    200 & 0\\ 
 \hline\hline 
\end{tabular} 
\caption{Values of $\frac{\overline g}{d}$ in the case of two levels 
with equal pair degeneracy $\Omega_1=\Omega_2=100$ for a few values of n.
}
\label{tab:3}
\end{center} 
\end{table}

We now abandon the assumption of s.p.l. of unit pair degeneracy and
address the question of the impact on $\frac{\overline g}{d}$
of the pair degeneracy of the s.p.l.. In the simple case of one 
pair in two levels the generalization of (\ref{eqc}) is easily found to read
\be\label{A1}
{\cal E}=(e_1+e_2)-g\frac{\Omega_1+\Omega_2}{2}-\sqrt{
\left[g\left(\frac{\Omega_1+\Omega_2}{2}\right)\right]^2+d^2+gd(\Omega_2-\Omega_1)
}
\ee
which yields 
\be\label{A2}
\frac{\overline g}{d}=2\left|\frac{\Omega_1-\Omega_2\pm 2i\sqrt{\Omega_1\Omega_2}}
{(\Omega_1+\Omega_2)^2}\right|=\frac{2}{\Omega_1+\Omega_2},
\ee
being $d=e_1-e_2$, from where it is seen that the singularity
moves away from the imaginary axis when $\Omega_1\neq\Omega_2$.

Sticking to the case $\Omega_1=\Omega_2=\Omega$ one gets from
(\ref{A2})
\be
\frac{\overline g}{d}=\left|\frac{i}{\Omega}\right|
\ee
and from (\ref{A1}) (choosing $e_1+e_2=0$) 
\be
{\cal E}=-g\Omega-\sqrt{d^2+g^2\Omega^2}
\ee
which coincides with (\ref{eqc}) when $\Omega=1$.
Thus the larger $\Omega$ is, the closer to the real axis the branch 
point is.
This occurrence has a bearance on the physics occurring on 
the real axis: in fact not only the energy but also the pair wave function
is strongly modified by $\Omega$, in the sense that the larger $\Omega$ is the sooner in $g$ the two components of \eqref{B} become equally weighted.

In the limit $\Omega\rightarrow\infty$ where the pair
becomes a true boson the singularity \emph{lies} on the real axis.

To assess the impact of the s.p.l. filling on $\overline g/d$ 
we quote in Table 3 the
numerical values of the latter
in the case of two levels with 
equal pair degeneracy $\Omega_1=\Omega_2=100$ for a few values of $n$.
From Table 3 one sees that $\overline g/d$ 
reaches its minimum at half filling being symmetric around the minimum. 

Finally we give a striking example illustrating the crucial role 
played by the s.p.l. degeneracy on $\overline g/d$. For this purpose 
we consider the two following cases 
\begin{itemize}
\item [a)] $L=2$, $\Omega_1=\Omega_2=100$, $\Omega_{\rm Tot}=200$, $d=1$, $n=100$
\item [and]
\item [b)] $L=200$, $\Omega_1=\Omega_2=\cdots=1$, $\Omega_{\rm Tot}=200$, $d=1/199$, 
$n=100$.
\end{itemize}

Then from the explicit computation
it turns out that in a)  $\overline g/d=0.0068$ 
(see Table 3) whereas in b) 
$\overline g/d=0.65$. Thus even in situations where analytical results 
are hard to achieve, the numerical analysis suggests that 
$\overline g/d\propto1/\tilde\Omega$, being $\tilde\Omega$
the degeneracy of each s.p.l..

In concluding this Section we return to the problem of the
thermodynamic limit previously mentioned pointing out that
two quite different
variational solutions, namely the Richardson \eqref{n} and the BCS \eqref{BCS} 
projected on a given particle number $M$ which reads 
\be
P_M|BCS>=\left(\sum_k\frac{v_k}{u_k}\hat c^\dagger_{\vec k\uparrow}
\hat c^\dagger_{-\vec k\downarrow}\right)^M,
\ee
yield the same ground state energy. 
Whether this remains true for other observables as well is an issue 
worth to be further explored.

\section{The spectrum of the pairing Hamiltonian}
In this Section we shortly examine the spectrum of the pairing Hamiltonian 
in the non-degenerate situation where (\ref{HP}) predicts new
excited states (referred to as Gaudin excitations) beyond those associated
with the breaking of pairs (referred to as seniority excitations), which are
the only ones present in the degenerate case.

These new kind of excited states correspond to the raising of pairs into
higher-lying s.p.l. and are classified in terms of the
Gaudin's number $N_G$~\cite{Gau95}, which yields the number of pair 
energies staying finite when $g$ goes to infinity because they remain
trapped in the grid of the s.p.l.. 
For fixed $N_G$ (of course $0\leq N_G\leq n$), the number of such excited states
is ${\Omega\choose N_G}-{\Omega\choose N_G-1}$.

\begin{figure}[t]
\vspace{-2cm}
\begin{minipage}[b]{0.5\textwidth}
\centering
\includegraphics[scale=0.8,clip,angle=0]{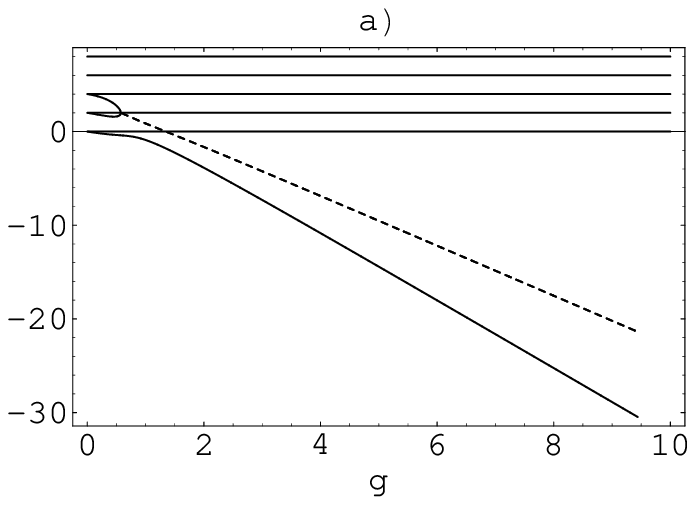}%
\end{minipage}
\begin{minipage}[b]{0.5\textwidth}
\centering
\includegraphics[scale=0.8,clip,angle=0]{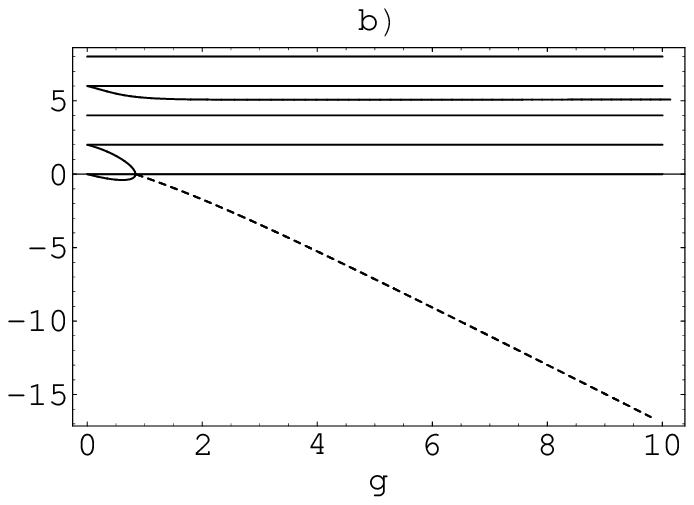}%
\end{minipage}
\begin{minipage}[b]{0.5\textwidth}
\centering
\includegraphics[scale=0.8,clip,angle=0]{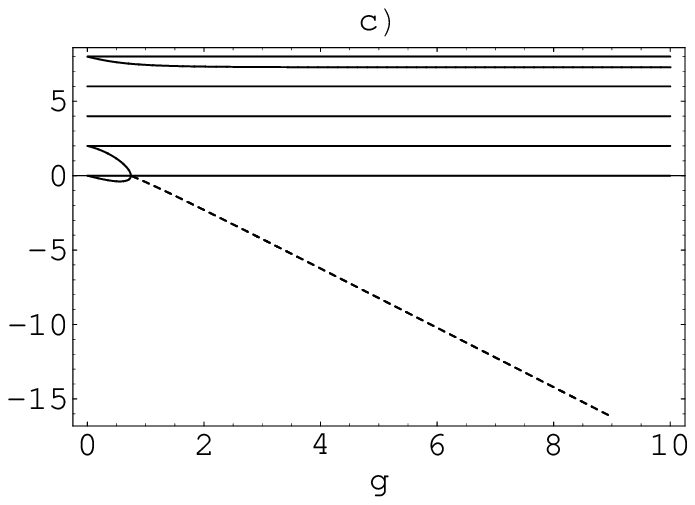}%
\end{minipage}
\begin{minipage}[b]{0.5\textwidth}
\centering
\includegraphics[scale=0.8,clip,angle=0]{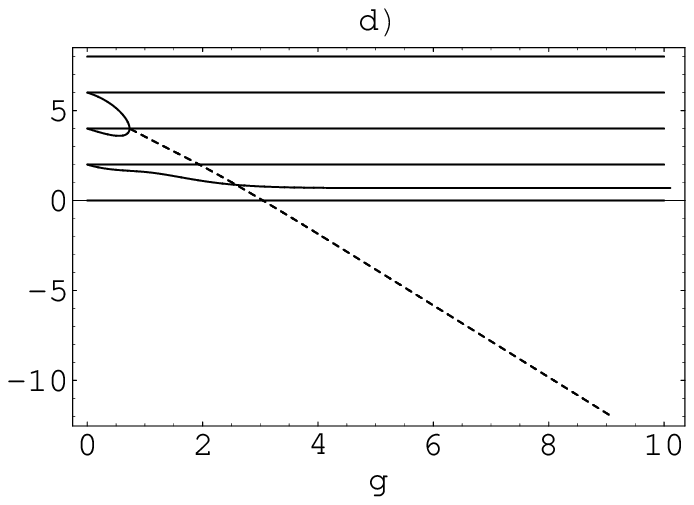}%
\end{minipage}
\begin{minipage}[b]{0.5\textwidth}
\centering
\includegraphics[scale=0.8,clip,angle=0]{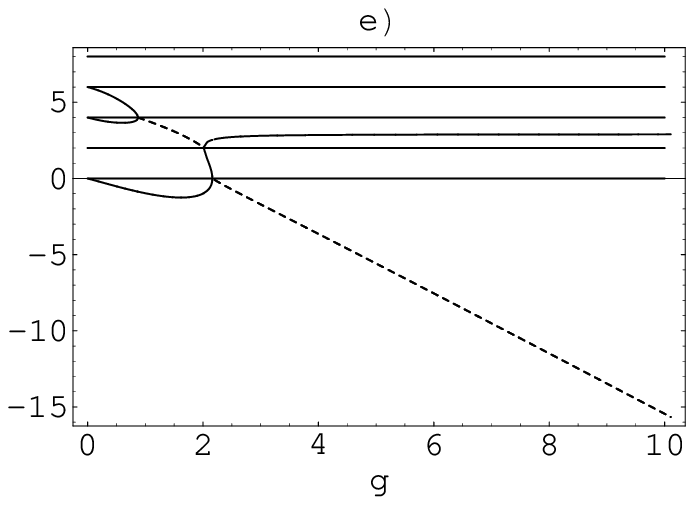}%
\end{minipage}
\begin{minipage}[b]{0.5\textwidth}
\centering
\includegraphics[scale=0.8,clip,angle=0]{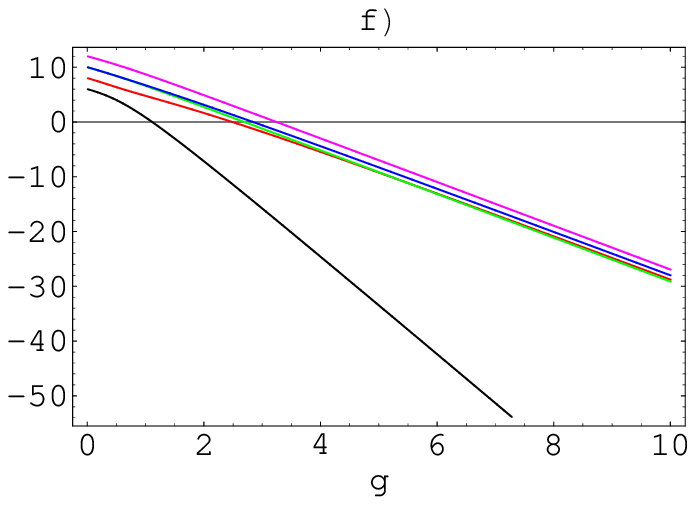}%
\end{minipage}
{\caption{
The real part of the exact solutions $E_\nu$ of the Richardson's equations (6) for a 
system of $n$=3 pairs living in a set of 
L=5 equispaced non-degenerate ($\Omega_k=1$) single particle levels with
energies $e_k=k$ ($k=0,\cdots 4$) are displayed versus 
the strength $g$. Panel $a$: ground state; 
Panels $b$-$e$: excited states with Gaudin's number $N_G=1$;
panel $f$: total energy of the ground state (lower
curve) and of the $N_G=1$ states (upper curves).
}\label{fig3}} 
\end{figure}

As an illustration of these excitations
we consider the simple example of 3 pairs in 5 equispaced levels of unit 
pair degeneracy. In this case there are
4 states with $N_G=1$ and 5 with $N_G=2$ (note that $N_G=3$ is not allowed).
In Fig.~\ref{fig3} we display the pair energies versus $g$ for the ground state 
(panel $a$) and for the $N_G=1$ states (panels $b$-$e$) together with 
the associated total energies (panel $f$). 
Note that while for the ground state and for each of
the first three excited states only one $g_{\rm crit}$ exists, for the fourth
$N_G=1$ excited state (panel $e$) a complicated escaping mechanism takes place,
giving rise to three $g_{\rm crit}$.
In the last panel, where the total energy of the above states is plotted, 
it clearly appears that, for $g$ high enough, the states tend to
group according to their Gaudin number in accord with the finding of~\cite{Yuz}.

Turning to 
the seniority excitations of \eqref{HP},
their number, for levels with unit pair degeneracies, is
${\Omega\choose 2s}{\Omega-2s\choose n-s}$ (of course it must be 
$0\leq s\leq\rm{min}\{n,\Omega-n\}$).

\begin{figure}[ht]
\begin{center}
\includegraphics[scale=0.75,clip,angle=0]{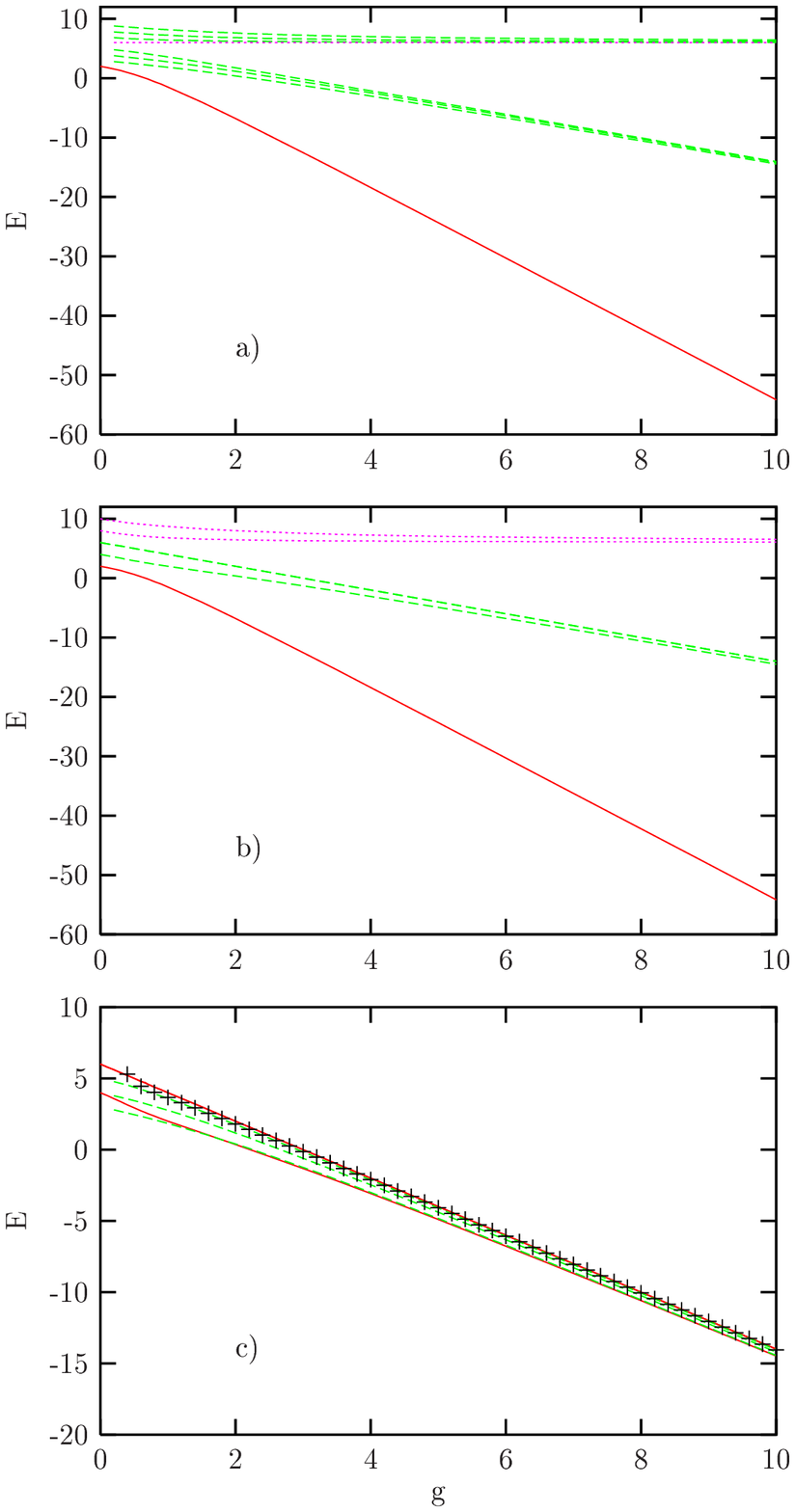}
\vspace{-4.cm}
{\caption{
a) Exact energies of the ground state (solid), 
seniority $s=1$ (dashed) and $s=2$ (dotted) states
for a system of 2 pairs in 4 equispaced single particle levels 
of energies
0, 1, 2 and 3,
respectively, and pair degeneracy 1. The 
$s=2$ state trivially corresponds to $E=e_1+e_2+e_3+e_4=6$.
b) Exact energies of the $s=0$ states with $N_G=0$ (solid), $N_G=1$ (dashed) 
and $N_G=2$ 
(dotted).
c) Comparison between the $N_G=1$,
$s=1$ states (dashed) and the $N_G=1$, $s=0$ (solid) excited states.
The crosses correspond to the approximate solution (37) for $N_G=1$.
}\label{fig4}} 
\end{center} 
\end{figure}

Again for illustration we consider the very simple set of configurations
associated with 2 pairs living in 4 equispaced single particle 
levels with unit energy distance and pair degeneracy. In this case 13
seniority excited states occur, 12 with $s=1$ and 1 
with $s=2$. 
Of the former, owing to the degeneracy
entailed by the symmetry of \eqref{HP}, only 6 show up in Fig.~\ref{fig4}a, 
where the $g$-behavior
of the seniority eigenvalues is displayed.
These can be analytically computed since they are obtained
as a solution of a two-levels problem, the other two levels being blocked
by the broken pair.  
Assuming, e.g., that the blocked single particle levels correspond to 
the energies
$e_1$ and $e_2$ (the other cases obtain with trivial permutations) they
read
\be
{\cal E}(n=2,s=1) = e_1+e_2+e_3+e_4-g
\pm\sqrt{g^2+(e_3-e_4)^2}~,
\ee
with obvious meaning of the symbols.
The $\pm$ sign accounts for the splitting of the $s=1$ states into 
two families. It turns out that the energy of the lower lying family 
comes pretty close to the energy of the $N_G=1$, $s=0$ state (compare
Fig.~\ref{fig4}b), whereas the energy of the higher lying family 
comes close to the one of the $N_G=2$ state.
Moreover these are almost degenerate with the $s=2$ state, corresponding
to a configuration where all the four levels are blocked: hence their
energy is simply given by the sum of the unperturbed single particle energies.

Generalizing the above example 
we argue that the excited states of the pairing 
Hamiltonian are identified by two quantum numbers,
namely $N_G$ and $s$. However the associated eigenvalues 
group, as $g$ becomes large, into families characterized 
by the number $N_B$ of pairs blind to the pairing interaction: whether 
they are so because they are broken or because they are trapped
it does not matter for $g$ sufficiently large.


For the energies of these states,
inspired by (\eqref{Es}), we accordingly surmise
the following approximate expression
\ba
{\cal E}(n,N_B)={\cal E}_{g.s.}(n)-{\cal E}_{g.s.}(N_B)+2 N_B {\bar e}~,
\label{EgsYs}
\ea
which we test against the exact solution in the $N_B=1$
case. Confining ourselves to \eqref{EgsY} the above becomes
\be
{\cal E}(n,N_B=1)-{\cal E}(n,N_B=0) =
2\left[
\frac{g\Omega}{2}+\frac{m^{(2)}}{(g\Omega/2)}-\frac{m^{(3)}}{(g\Omega/2)^2}
+\frac{m^{(4)}-2\left[m^{(2)}\right]^2}{(g\Omega/2)^3}
\right]
\label{Es1}
\ee
whose first term,
significantly, is equal to $2\Delta$.

Notwithstanding the shortcomings of formula (\ref{EgsYs})
yet we have found, as seen in Fig.~\ref{fig4}c, that
it provides a good account for the exact results, except at low $g$
namely  in the domain of the critical values of the coupling
(in our example these occur at 0.66, 1, 2 and 3 in 
units of the spacing between the single particle levels).
In addition (\ref{Es1}) shows that to leading order, which is the accuracy
of formula (\ref{Es1}), the Gaudin and the seniority 
excitations
are both characterized by the same gap $\Omega g$. 

\section{Conclusions}
The problem of the pairing Hamiltonian, extensively treated in a number of papers,
is still the object of many investigations. 
Such an Hamiltonian employs a drastically simplified interaction
and, as a consequence, cannot be viewed as a realistic one. 
It yields, for example, a gap
too large by a factor $(4e)^{1/3}$ with respect 
to the accurate predictions of Gorkov ~\cite{Gorkov} and Heiselberg 
~\cite{Heiselberg}. Yet the fact of being exactly solvable 
makes \eqref{HP} interesting 
and helpful for the researches in a variety of fields. To mention a couple 
of these, we remind the microscopic 
derivation of the Arima-Iachello model and the physics of small 
superconducting metallic grains, which has lately attracted much interest.

In this work we have addressed a few aspects of the problem which, in our 
view, still deserve further studies. First we have proved in the path integral 
framework and in the degenerate case the superfluidity of the pairing 
Hamiltonian solution in a ``finite'' system. This item was already lately 
treated by us ~\cite{Barbaro:2004nk}. 
Here, however, we have added the important 
finding that the saddle point of the effective action derived in 
~\cite{Barbaro:2004nk} is 
exactly fixed by the BCS gap.

Then we have compared the $g$-behavior of the ground state energy in the 
degenerate case with the situation when many single particle levels are 
active. Although this comparison has been limited to levels with unit pair 
degeneracy it has allowed us to grasp the impact of the s.p.l. 
distribution on the system's energy and to get, at least in the half filling
situation, an orientation on
the regimes in which the system lives.

Next we have derived an expansion of the system ground state energy 
${\cal E}_{g.s.}(g)$ in the inverse of the coupling constant 
(strong coupling expansion). The formula we have obtained generalizes the one  
of Ref.~\cite{Yuz} since it is expressed in terms of the statistical moments of the 
s.p.l. distribution and is thus valid for \emph{any} distribution.
This appears to be useful for some recent researches on the density functional 
approach to the pairing Hamiltonian \cite{Papenbrock}, but it is also helpful
for a deeper understanding of the dynamics of a system ruled by (\ref{HP}).
Indeed the investigation of the domain of validity of the expansion has lead us
to conclude that such a domain is set by a singularity of the function ${\cal E}(g)$
located somewhere in the complex 
$g$-plane. Although we have not been able to find a general expression for this 
singularity (indeed it depends from the specificity of each case)
we have numerically found where its modulus $\overline g$ lies 
at least for s.p.l. with unit pair degeneracy
and how its location is affected by the distribution of the levels 
(in particular by their degeneracy)
defining the space on which the Hamiltonian (\ref{HP}) acts.
This is of relevance because we conjecture that
the physics is more or less affected by the singularity 
depending upon its proximity to the positive real axis, as it has been
hinted by the simple $n=1$, $L=2$ case: accordingly we surmise that
if the singularity is located 
remotely from the real axis one deals with a system which is predominantly 
fermionic, if it is closer to the real axis (as it may be the case when the 
s.p.l. have large pair degeneracies)
one deals with a system in which a process of bosonization is taking place.

Finally we have discussed the excitation spectrum of the pairing Hamiltonian. 
We have found that the Gaudin and seniority excitations are close to each 
other, at least for small $s$ and $N_G$. We have proposed an approximate analytic 
expression for these excitations that appears to account satisfactorily
for the exact result. Importantly, both the Gaudin and the seniority excitations 
display the same gap in the leading order expressed by our approximation.

\end{document}